\documentstyle[12pt,preprint,aps,floats,tighten,psfigure,prabib]{revtex}
\begin{document}
\flushbottom
\pagestyle{plain}
\pagenumbering{arabic}
\textwidth 6in
\textheight 9in
\parskip=12pt
 
%
\def\snw{\sin^2\theta_W}
\def\deg{$^\circ$}
\def\etal{{\it et al.}}
\def\inch{$\,^{\prime\prime}$}			
\def\gtorder{\mathrel{\raise.3ex\hbox{$>$}\mkern-14mu
             \lower0.6ex\hbox{$\sim$}}}
\def\ltorder{\mathrel{\raise.3ex\hbox{$<$}\mkern-14mu
             \lower0.6ex\hbox{$\sim$}}}
%

\title{ \bf
SAMPLE: PARITY VIOLATING ELECTRON SCATTERING \\
FROM HYDROGEN AND DEUTERIUM\\ }

\author{
E.J.~Beise,$^3$\,
J.~Arrington,$^1$,
D.~H.~Beck,$^2$\,
E.~Candell,$^4$\,
R.~Carr,$^1$\,
G.~Dodson,$^5$\,
K.~Dow,$^5$\,
F.~Duncan,$^3$\,
M.~Farkhondeh,$^5$\,
B.W.~Filippone,$^1$\,
T.~Forest,$^2$\,
H.~Gao,$^2$\,
W.~Korsch,$^1$\,
S.~Kowalski,$^5$\,
A.~Lung,$^{1,3}$\,
R.~D.~McKeown,$^1$\, 
R.~Mohring,$^3$\,
B.~A.~Mueller,$^1$\,
J.~Napolitano,$^4$\,
M.~Pitt,$^1$\,
N.~Simicevic,$^2$\,
E.~Tsentalovich,$^5$\,
S.~Wells$^5$}

\address{
$^1$\ California Institute of Technology, Pasadena, California 91125\\
$^2$\ University of Illinois at Urbana-Champaign, Urbana, Illinois 61801\\
$^3$\ University of Maryland, College Park, Maryland 20742\\
$^4$\ Rensselaer Polytechnic Institute, Troy, New York 12180\\
$^5$\ MIT-Bates Linear Accelerator Center, Middleton, Massachusetts, 01949\\
}

\maketitle
 
\begin{abstract}

Recently, there has been considerable theoretical interest
in determining strange quark contributions to hadronic matrix 
elements. Such matrix elements can be accessed through the nucleon's
neutral weak form factors as determined in
parity violating electron scattering. The SAMPLE experiment
will measure the strange magnetic form factor $G_M^s$ at low
momentum transfer. By combining measurements from hydrogen and
deuterium the theoretical uncertainties in the measurement
can be greatly reduced and the result will be limited by 
experimental errors only. A summary of recent progress on
the SAMPLE experiment is presented.

\end{abstract}

\section*{Introduction and Background}

Elastic electron scattering has successfully been used as a probe of nucleon
structure for many years. The electromagnetic properties of the proton
are now very well known\cite{GEP}, and recently there has been considerable
progress in measuring the electric and magnetic form factors of the
neutron \cite{GEN,GMN}.  
Additional and complementary information on nucleon structure can
be obtained through the use of neutral weak probes. For example,
there has been much recent theoretical interest in the possibility
that sizeable strange quark contributions to nucleon matrix elements
may exist. The two most cited pieces of experimental evidence 
are measurements of the $\pi$-nucleon $\Sigma$ term, from which the 
scalar matrix element $\langle N\vert\overline{s}s\vert N\rangle$ 
can be obtained \cite{Gasser}, and 
measurements of the nucleon's spin-dependent structure functions in 
deep-inelastic lepton scattering\cite{gspin}, from which the axial 
current $\overline{s}\gamma_{\mu}\gamma_5 s$ is extracted. In each
case the $s$-quark contribution to the proton is about 10-15$\%$,
although both results are sensitive to theoretical interpretation.

Parity violating electron scattering provides the opportunity to
investigate the vector matrix element 
$\langle N\vert\overline{s}\gamma_\mu s\vert N\rangle$.
This matrix element can be determined directly from a measurement
of the neutral weak form factor of the proton, {\it i.e.}, the
interaction between a proton and electron through the exchange
of a $Z$ boson. The electromagnetic and weak form factors 
can be constructed as a sum of individual quark distribution 
functions multiplied by coupling constants given by the Standard 
Model for Electroweak Interactions. The lepton currents are
completely determined, and the hadronic currents are the
information to be extracted by experiment. The electromagnetic
coupling gives the well known Sachs form factors $G^{p,n}_{E,M}$.
Neglecting quarks
heavier than the $s$-quark and making the assumption that the
proton and neutron differ only by the interchange of $u$ and $d$
quarks, the neutral weak vector form factors can be expressed in terms
of the EM form factors in the following way:
\begin{equation}
G^{Z,p}_{E,M} = \left({1\over 4} - \sin^2\theta_W\right)
  \left[1+R^p_V\right]G^p_{E,M} - {1\over 4}\left[1+R^n_V\right]G^n_{E,M}
   - {1\over 4}\left[1+R^s_V\right]G^s_{E,M}
\end{equation}
\begin{equation}
G^{Z,n}_{E,M} = \left({1\over 4} - \sin^2\theta_W\right)
  \left[1+R^p_V\right]G^n_{E,M} - {1\over 4}\left[1+R^n_V\right]G^p_{E,M}
   - {1\over 4}\left[1+R^s_V\right]G^s_{E,M}
\end{equation}
The factors $R^i_V$ are weak radiative corrections which must
be applied to account for higher order processes.
In addition there is an axial vector coupling which leads to
\begin{equation}
G^Z_A = -{1\over 2}\left[1+R_A^{T=1}\right]g_A\tau_3 
      + {\sqrt{3}\over 2}R_A^{T=0}G^{(8)}_A 
      + {1\over 4}\left[1+R^s_A\right]G^s_A
\end{equation}
where $\tau_3$=+1(-1) for the proton(neutron).

The term involving the SU(3) isoscalar form factor $G^{(8)}_A$ is generally
ignored since it is not present at tree level, and an estimate of 
$R^{T=0}_A$~\cite{PhysRep} that it is in fact suppressed relative to the 
dominant first term.
The isovector axial form factor $g_A(0)$=1.26 is determined from neutron
beta decay.  With the exception of $G^n_E$, the electromagnetic 
form factors are determined with good precision. The axial 
strange form factor $G^s_A$ is the same
quantity extracted from polarized deep inelastic lepton scattering.
The only undetermined
quantities are the strange quark contributions $G^s_{E,M}$. 
At $Q^2$=0, $G^s_E$=0 because the proton has no net strangeness. 
The magnetic form factor $G^s_M(0)$ is not well constrained, and 
the $Q^2$ dependence of all three strange form factors is 
unknown. This has stimulated a program of parity violation
experiments at Bates~\cite{BMck,Pitt},  
CEBAF~\cite{Beck,Souder,Beise} and Mainz~\cite{Mainz_PV}.

Since these experiments were proposed several years ago,
many theoretical predictions have been put forth to estimate
the size of the $s$-quark contributions. The vector
contributions are characterized by 
two parameters: the ``strange magnetic moment''
$\mu_s = G^s_M(0)$, and the ``strangeness radius''
$r^2_s = -{1\over 6}{d F_1^s\over d Q^2}$. 
Predictions of the various models, which range from
vector-meson dominance-like mechanisms to Skyrme calculations
to those involving kaon loop diagrams, were described in
detail in references~\cite{PhysRep} and \cite{Forkel}
and an updated list is shown in table I. A notable point 
is that although different models have similar predictions
for the magnitude of $\mu_s$, (with a few exceptions) the 
estimates of $r^2_s$ vary widely depending on the type of mechanism assumed.

\goodbreak
\begin{center}
\begin{table}
\caption[Table I]{Theoretical predictions for $G^s_M(0)$
and $r^2_s$, after Musolf {\it et al.}~\cite{PhysRep} with
recent additions.}
\begin{tabular}{lll}
\tableline
Type of Calculation (reference) & $\mu_s$  & $r^2_s$ (fm$^2$)  \\
\tableline
Poles~\cite{Jaffe89} & $-0.31\pm 0.009$ & $0.14\pm 0.07$  \\
Poles~\cite{Hammer}  & $-0.24\pm 0.03$ & $0.21\pm 0.03$ \\
Kaon Loops~\cite{MusBur} & $-0.40\longrightarrow -0.31$ & $-0.03\pm 0.003$ \\
Kaon Loops~\cite{Koepf} & $-0.03$ & $-0.01$  \\
``Loops and Poles''~\cite{Cohen} & $-0.28\pm 0.04$  & $-0.03\pm 0.01$ \\
SU(3) Skyrme~\cite{Park91} & $-0.33\longrightarrow -0.13$ 
   & $-0.11\longrightarrow -0.19$  \\
SU(3) chiral hyperbag~\cite{Hong} & $0.42\pm 0.30$ &  \\
SU(3) chiral color dielectric~\cite{Phatak} 
   & $-0.40 \longrightarrow -0.03$ & $-0.003\pm 0.002$ \\
Chiral quark-soliton~\cite{Kim} & $-0.45$ & $-0.17$ \\
Constituent Quark~\cite{Ito} & $-0.13\pm 0.01$  & $-0.002$ \\
``QCD Equalities''~\cite{Lein} & $-0.73\pm0.30$ &  \\
\tableline
\end{tabular}
\label{table1}
\end{table}
\end{center}

\section*{Parity-Violating Electron Scattering at MIT-Bates}

Parity violating electron scattering is evaluated in terms of the asymmetry
in the cross section for the scattering of right- and left-helicity 
electrons from an unpolarized target.
For elastic scattering from a free nucleon, the asymmetry can be written
as a sum of three terms which reflect the interference between the NW
and EM interactions:
\begin{equation}
A_p = \left[{G_F Q^2\over \sigma_p\pi\alpha\sqrt{2}}\right]
   \left[ \varepsilon G_E^p G_E^{Z,p}   + \tau G_M^p G_M^{Z,n} 
   - {1\over 2}\left(1-4\snw\right)\varepsilon^{\prime} G_M^p G_A^{Z,p}\right]
    \, , 
\end{equation}
\begin{equation}
A_n = \left[{G_F Q^2\over \sigma_n\pi\alpha\sqrt{2}}\right]
   \left[ \varepsilon G_E^n G_E^{Z,n}   + \tau G_M^n G_M^{Z,p} 
   - {1\over 2}\left(1-4\snw\right)\varepsilon^{\prime} G_M^n G_A^{Z,n}\right]
    \, , 
\end{equation}
where $\tau = Q^2/4M_p^2$, 
$\varepsilon = \left(1+2(1+\tau)\tan^2{\theta\over 2}\right)^{-1}$ 
and 
$\varepsilon^\prime = 
\sqrt{\left(1-\varepsilon^2\right)\tau\left(1+\tau\right)}$.
The terms $\sigma_{p,n} = \varepsilon\left(G^{p,n}_E\right)^2 + 
\tau\left(G^{p,n}_M\right)^2$ are proportional
to the elementary unpolarized cross sections.
The kinematic factors result in the first two terms of 
the asymmetry dominating at forward angles and the latter
two terms contributing at backward angles, although the term containing the 
axial vector form factor $G_A^Z$ is suppressed by the factor 
$(1-4\snw)$. 

The SAMPLE experiment at Bates~\cite{BMck}
is a measurement of $A_p$ at backward
angles ($130^\circ < \theta < 170^\circ$) and $E_{lab}$ = 200 MeV, 
resulting in $Q^2 = 0.1$ (GeV/c)$^2$. At these kinematics the parity 
violating asymmetry for a proton target is $-7.3\times 10^{-6}$, 
assuming no contribution from $s$-quarks, and is dominated by the 
middle term and thus is sensitive to $G^s_M(0)$. 
The SAMPLE goal is to achieve a
statistical error of approximately 7\%. The dominant 
experimental systematic error
is expected to come from knowledge of the beam polarization.
Recently, Garvey {\it et al.}~reanalyzed   quasielastic 
$\nu$($\overline\nu$)-$p$ scattering data from BNL
experiment E734~\cite{Garvey}, and extracted a value of
$G^s_M(0) = -0.4\pm 0.7$ with some model dependence, extrapolating
the result from  $Q^2\sim 0.75$ (GeV/c)$^2$. 
The SAMPLE experiment would improve this result by a factor
of 3, to an absolute error of $\delta G^s_M = 0.22$. 

\subsection*{Update on the SAMPLE Experiment}

\begin{figure}[htb]
\centerline{
\psfigure[3.5in]{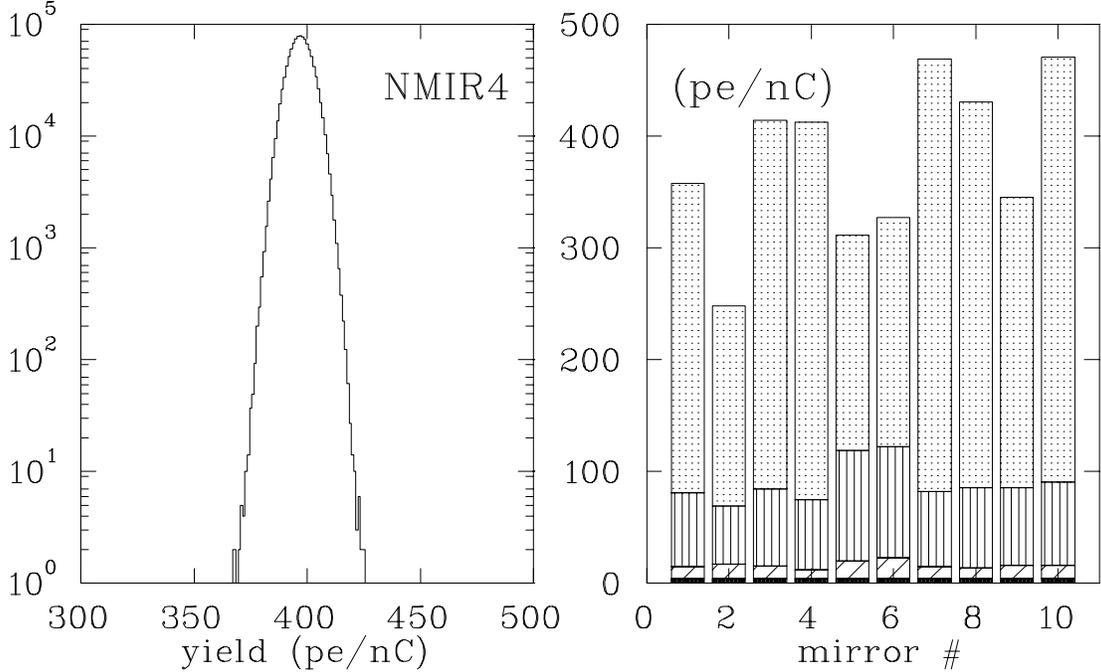} }
\vskip 0.2in
\caption{(a) Representative histogram showing normalized 
yield in one phototube.
(b) Bar plot showing contributions from various sources to normalized yield.
The different contributions are: light yield from
hydrogen (dots), background from hydrogen (vertical lines), 
background from the target cell (slanted lines), and from 
the beam line (solid).}
\label{fig:nyield}
\end{figure}

In the SAMPLE experiment, 
polarized electrons are incident on a 40~cm long liquid hydrogen
target. Elastically scattered electrons are detected in 
the backward direction by a large solid angle air
{\v C}erenkov detector consisting of ten mirrors which image the
target onto ten 8~inch photomultiplier tubes. The average current
of the beam is typically 40~$\mu$A, delivered 
with a 1\% duty cycle at 600~Hz.
The counting rate in each photomultiplier tube is very high, so
individually scattered electrons are not detected but the signal
is integrated over the 15~$\mu$sec beam pulse and normalized
to the charge in each burst. Background is measured independently
with empty target runs and by closing shutters in front of the phototubes.
In figure~\ref{fig:nyield} is shown a typical normalized 
yield distribution and
a bar plot representing the contributions of the light yield from
hydrogen (dots), the background from hydrogen (vertical lines), and 
background from the target cell (slanted lines), and the beam line (solid).

In September 1995 the first substantial data taking with 
polarized beam was carried out. Approximately 76 hours of
high quality asymmetry data were acquired, corresponding 
to a statistical error of 0.5~ppm. Analysis of the 
data is in progress to understand
and minimize the most important systematic errors before
beginning a long experimental run in spring 1996. Some features
of the data are discussed below.

\begin{figure}[htb]		%
\centerline{
\psfigure[3.5in]{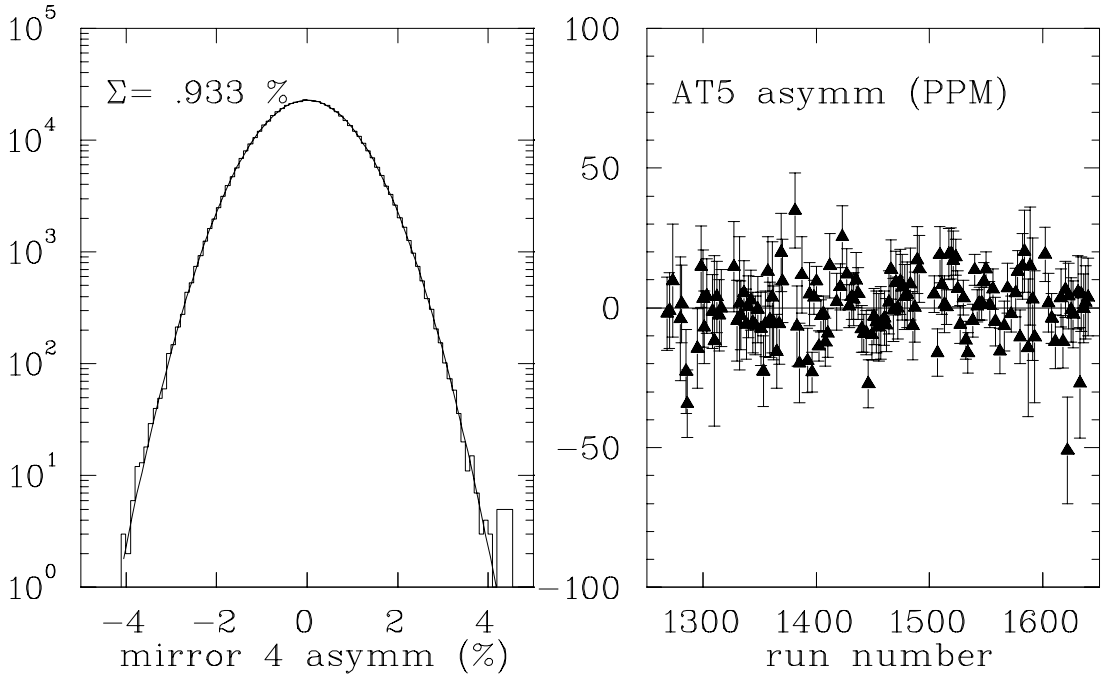} }
\vskip 0.2in
\caption{(a) Representative histogram of pulse-pair asymmetries
in one mirror for one 1/2 hour run. (b) Beam charge asymmetry at the
end of the Bates accelerator for the duration of the fall 1995 run.} 
\label{fig:asymm}
\end{figure}

The polarized beam is delivered to the SAMPLE apparatus at a maximum
repetition rate of 600~Hz. Circularly polarized laser light from
a Titanium-sapphire laser is incident upon a crystal made of bulk 
GaAs, from which an electron beam of approximately 40\% polarization 
is extracted. The circularly polarized light is generated by 
a linear polarizer followed by a Pockels cell which acts as a 
quarter wave plate when the appropriate voltage is applied.
The helicity of the electron beam is flipped by reversing the
polarity of the voltage on the Pockels cell. In addition,
a linear polarizer plate can be manually rotated in the laser beam
to reverse the polarity independently of the Pockels cell.
The beam control and data acquisition
are based upon the system used in previous parity violation
measurements at Bates \cite{C12}, where a systematic error 
$2\times 10^{-8}$ was reported. The beam helicity is chosen randomly
on a pulse-by-pulse basis, except that pulse pairs 1/60 sec
apart have opposite helicity. ``Pulse-pair'' asymmetries are formed
every 1/30 sec, greatly reducing sensitivity to 60~Hz electronic
noise and to drifts in beam properties such as current, 
energy, position and angle, which generally occur on time scales much
longer than 1/30 sec. In each 1/30 sec measurement period, there are
ten calculated asymmetries (in each of the ten ``time-slots'' of the 600 Hz
beam for each of the ten mirror signals, so that a
half-hour run contributes $5\times 10^{6}$ separate measurements of a 
pulse-pair asymmetry.
Shown in figure~\ref{fig:asymm}(a) is a typical distribution of
asymmetries measurements for one mirror in a half-hour run.
The width is about 1\%, consistent with the expected counting 
statistics per beam pulse.

The electron beam deposits approximately 550 watts of power into 
the target. Density fluctuations are minimized by subcooling
the liquid below its boiling point by a few degrees and by
rapidly circulating the fluid in a closed loop such that a 
packet of hydrogen is in the path of the beam for only a short
time. No density reduction was seen in the normalized yield at 
the level of a few percent as the beam current was varied
between 4 to 40~$\mu$A. Fluctuations in the normalized yield due to
variations in beam properties limit the accuracy with which 
this observable can be used to determine density changes. A
more sensitive determination is to monitor the width of
the pulse-pair asymmetry. In this observable no
change in density was seen at the level of 1\% or better.

Helicity correlations in the beam properties can cause
parity-conserving asymmetry contributions to the data.
The raw measured asymmetry $A_{raw}$ must be corrected for these effects.
The corrected asymmetry $A_c$ can be expressed as
\begin{equation}
A_c = A_{raw} - {1\over S} \Sigma_i {\partial S\over\partial\alpha_i}
  \delta\alpha_i^{LR} \, ,
\end{equation}
where $S$ is the normalized detector 
yield, $\alpha_i$ is one of
five beam parameters (position and angle in $x$ and $y$, and energy),
and $\delta\alpha_i^{LR} = \alpha_i^R - \alpha_i^L$ is the helicity
correlated difference in the beam parameter. 
To first order, the experiment is designed to be as insensitive
as possible to fluctuations in beam properties. This includes
feedback loops to stabilize the beam energy and position on target.
The corrections are then made with measured helicity correlations in 
the beam and the measured detector sensitivity to beam properties. 
The latter is performed by systematically changing the position and angle
of the beam on target with a system to rapidly control two sets of
steering coils and by changing the 
energy of the beam rapidly with an RF phase-shifting apparatus.
The sensitivity of the asymmetry to beam position differences
was typically 1\% per mm.  
The coil-pulsing system also allows us to minimize
the background in the detector as a function of beam location.
In the experiment of reference~\cite{C12} it was determined that
the largest source of beam fluctuations originated in the transport
of the laser beam. A feedback loop between beam charge asymmetry
and Pockels cell voltage minimizes helicity correlations due to
the laser beam. Figure~\ref{fig:asymm}(b) shows the beam charge asymmetry
at the exit of the accelerator. 

Analysis of the data is presently underway. Asymmetries on
the order of 100~ppm or larger
in the beam charge were seen and found to be due 
to problems with the laser feedback system, which were repaired part 
way through the run. These data have been removed from 
figure~\ref{fig:asymm}(b).
Helicity correlated shifts in  the
beam position at the target of 200~nm (vertical) and
200-600~nm (horizontal)
were also found throughout the run and are under investigation.
The feedback loop on beam position on target was commissioned,
and an additional feedback loop on beam energy was developed and 
will be tested in the upcoming run.

\subsection*{Deuterium}

At the SAMPLE kinematics the contribution from the axial vector
term $G_M G_A^{Z}$ is about 20\%. The weak radiative correction
to this term $R^{T=1}_A$ has been estimated to be $-0.34\pm 0.28$
\cite{MusHol}. This leads to a theoretical limit on the
ultimate uncertainty in the experimental result corresponding
to $\delta G^s_M \sim\pm 0.12$. A direct measure
of this radiative correction would therefore be useful.
In quasielastic scattering from deuterium at the same kinematics,
the axial term contributes approximately the same amount to the
asymmetry as in the proton, but the contribution from the
term proportional the $G^s_M$ is greatly reduced because the
proton and neutron contributions add incoherently and nearly
cancel. Taking the ratio $A_p/A_d$ would as a result reduce
the theoretical error to $\delta G^s_M\sim\pm 0.01$. 
The ratio $A_p/A_d$ would be insensitive to systematic errors
in the measured beam polarization. At the SAMPLE
kinematics the expected value of the deuteron asymmetry is
$A_d = -9.6\times 10^{-6}$ and the counting rate is about
twice that for the proton. The only required modification to the
SAMPLE apparatus is a recovery
system for the costly deuterium. In 1994 an additional 1000 hours 
of beam time was approved for a deuteron measurement~\cite{Pitt}.

In the ``static'' approximation, the deuteron asymmetry can be 
written as an incoherent sum of contributions from the proton 
and neutron weighted by the unpolarized cross sections:
\begin{equation}
A_d = {\sigma_p A_p + \sigma_n A_n\over\sigma_d} \, .
\end{equation}
Hadjimichael, Poulis and Donnelly investigated the dependence
of $A_d$ on the structure of the deuteron \cite{Hadji}
and found that is insensitive to corrections to the static model at the
level of 1-2\%.

Because the SAMPLE detector has no energy resolution, the
measured deuteron asymmetry includes contributions from
elastic $e$-$d$ scattering and from electrodisintegration.
The elastic asymmetry $A_{el}$ at backward angles was estimated by 
Pollock~\cite{Pollock} to be very sensitive to $G^s_M$.
The electrodisintegration asymmetry $A_{ted}$, as well as
other small contributions from elastic scattering,
was calculated by Musolf and Donnelly~\cite{MusDon}.
Fortunately, all of these contributions are small because
the scattering is greatly dominated by the quasielastic
cross section. The resulting contribution to the measured
asymmetry can be summarized as:
\begin{eqnarray}
A_{tot}&= {1\over \sigma_{tot}} \left[A_{QE}\sigma_{QE} + 
   A_{el}\sigma_{el} + A_{ted}\sigma_{ted}\right] \\
   &= (-9.6)\left({47\over 52}\right) + (10.3)\left({1\over 52}\right)
    + (-12.8)\left({4\over 52}\right) \\
   &= -9.5\, {\rm ppm} \, .
\end{eqnarray}

\begin{figure}[htb]		%
\centerline{
\psfigure[3.5in]{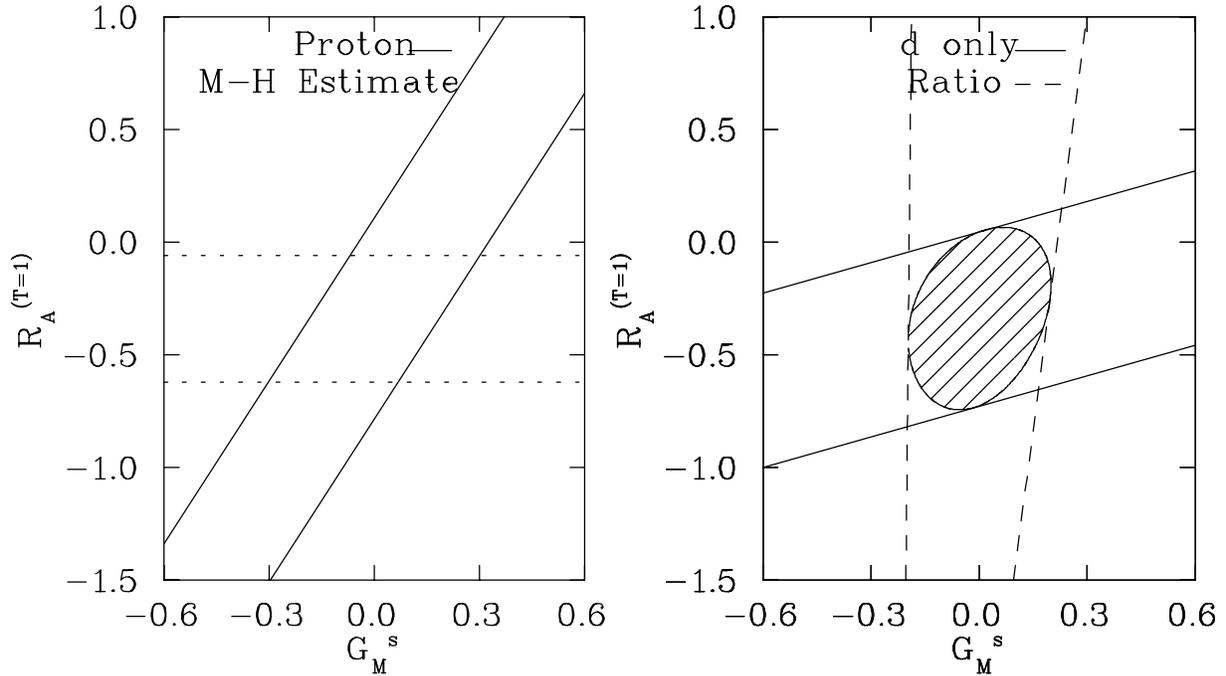} }
\vskip 0.2in
\caption{Left Panel: expected 1$\sigma$ limits on $G_M^s$ 
from hydrogen measurement
alone. Right Panel: $\sigma$ limits on $R^1_A$ and $G_M^s$ from a
combined measurement of hydrogen and deuterium. }
\label{fig:ra_vs_gm}
\end{figure}

The degree to which the deuteron measurement will improve the
determination of $G^s_M(0)$ can be summarized in 
figure~\ref{fig:ra_vs_gm}. The solid lines
in the left-hand graph show the constraints on $G^s_M(0)$ 
that would result from the $A_p$ alone. The result is
clearly correlated with $R_A^{T=1}$. The dashed lines
represent the calculation of $R_A^{T=1}$ by 
Musolf and Holstein along with the estimated uncertainty.
In the right panel is shown the sensitivity of a measurement
of $A_d$ (solid lines) which is nearly independent
of $G^s_M$, and the ratio $A_p/A_d$ (dashed lines),
which is nearly independent of $R^{T=1}_A$. 
The ellipse represents the resulting constraints.

\section*{Conclusions}

Theoretical interest in understanding the role of strange quarks in 
the nucleon has stimulated a new generation of experiments
in parity-violating electron scattering and also in
neutrino scattering~\cite{LSND}. Currently very 
little information is known. Considerable progress
was made in 1995 on the SAMPLE experiment at Bates,
and a long data-taking run on the proton is expected to occur in
1996. Additional running with a deuterium target has
also been approved.

This work was supported by NSF grants PHY-9457906/PHY-9229690 (Maryland), 
PHY-9420470 (Caltech), PHY-9420787 (Illinois), 
PHY-9208119 (RPI) and DOE cooperative agreement 
DE-FC02-94ER40818 (MIT-Bates).

\end{document}